# The Search for Extrasolar Earth-like Planets

S. Seager

*Department of Terrestrial Magnetism, Carnegie Institution of Washington, Washington, DC, USA and Institute for Advanced Study, Princeton, NJ, USA*

___

**Abstract**

The search for extrasolar Earth-like planets is underway. Over 100 extrasolar giant planets are known to orbit nearby sun-like stars, including several in multiple-planet systems. These planetary systems are stepping stones for the search for Earth-like planets; the technology development, observational strategies, and science results can all be applied to Earth-like planets. Stars much less massive than the sun—the most common stars in our Galaxy—are being monitored for the gravitational influence of Earth-like planets. Although Earth-like planets orbiting sun-like stars are much more difficult to detect, space missions are being built to detect them indirectly due to their effects on the parent star and to quantify fundamental factors such as terrestrial planet frequency, size distribution, and mass distribution. Extremely ambitious space programs are being developed to *directly* detect Earth-like planets orbiting sun-like stars, and must tackle the immense technological challenge of blocking out the light of the parent star, which is brighter than the planet by six to ten orders of magnitude. Direct detection of radiation from the planet is necessary for the definitive goal of the search for Earth-like planets: the study of atmospheric spectral signatures for signs of severe disequilibrium chemistry that could be indicative of biological activity. In addition to technological development, a growing flurry of scientific activity has begun to: understand terrestrial planet formation and terrestrial planet frequency; model terrestrial-like planet atmospheres and evolution; articulate the biological signatures of our own Earth; and even to study Earth as an extrasolar planet by observation and analysis of the spatially unresolved Earth.

___

**Introduction**

For thousands of years people have wondered, are we alone? Is there life elsewhere in the Universe? For the first time in human history we are on the verge of being able to answer this question by detecting extrasolar Earth-like planets. A prime goal is to analyze the spectra of extrasolar Earth-like planet atmospheres in great detail to search for signs of life or habitability.

The solar system contains a broad diversity of planets (see Figure 1). Despite their very different appearance, the solar system planets are usually divided into two main categories: terrestrial planets and giant planets. The terrestrial planets include the four inner planets, Mercury, Venus, Earth, and Mars. The terrestrial planets are predominantly composed of rock and metals and have thin atmospheres. Atmosheric evolution, from both atmospheric escape of light gases and gas-surface reactions, has substantially changed each of the terrestrial planet atmospheres from their primitive state. Here I further define *extrasolar Earth-like planet* to mean a planet similar in mass, size, and temperature to Earth. I use the term *Earth twin* to refer to an Earth-like planet with liquid water oceans and continental land masses. The giant planets include the four outer planets, Jupiter, Saturn, Uranus and Neptune. These planets are vastly different from the terrestrial planets, with no rocky surfaces and masses hundreds of times those of the terrestrial planets. The giant planets are composed almost entirely of hydrogen and helium with massive atmospheres and liquid interiors. The giant planet atmospheres are primitive—little atmospheric evolution has taken place so that they contain roughly the same atmospheric gases as at their formation. The term *extrasolar giant planets* refers to planets with masses and sizes similar to Jupiter.

Direct detection of extrasolar planets is not possible with current telescopes and instrumentation. Earth-like planets in our "local neighborhood" of 300 light years are not intrinsically faint from an astronomical perspective—they are not fainter than the faintest galaxies ever observed (by the Hubble Space Telescope). Earth-like planets, however, pose a special challenge for observational detection and study due to the close proximity of the parent stars, which are orders of magnitude brighter and more massive than the planets (see Figure 2).

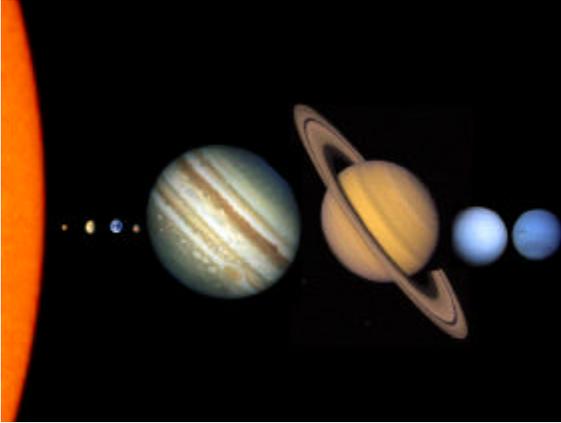

*Fig. 1 A collage of the solar system planets. Images from NASA. The planet sizes are approximately to scale but the planet separations are not. The four inner terrestrial planets are much smaller, less massive and therefore more difficult to detect in an extrasolar planetary system than giant planets like Jupiter and Saturn. Many decades will pass before we are able to obtain similar, spatially resolved images of extrasolar planets.*

Nevertheless the first step towards directly detecting extrasolar Earth-like planets has been achieved with the indirect detection of over 100 extrasolar giant planets [1,2,3]. The giant planets' gravitational influence on the parent stars can be detected. Furthermore, technological development is underway to detect Earth-like planets orbiting sun-like stars. Both the European Space Agency (ESA) and the US National Aeronautics and Space Administration (NASA) have ambitious extrasolar planet programs. These programs are starting with ground-based technological development and they are expected to culminate in large, billion dollar space missions.

The detection and characterization of extrasolar planets is a rapidly moving field with frequent new developments and huge promise for the future. This paper reviews the latest developments in the search for extrasolar Earth-like planets including the current status, relevant recent advances, and future plans.

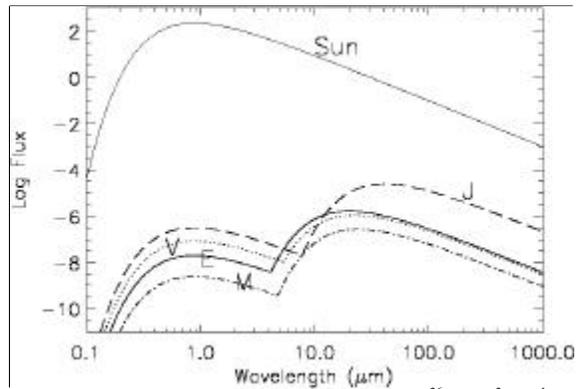

*Fig. 2 The approximate spectra (in units of $10^{-26}$ W m$^2$ Hz$^{-1}$) of some solar system bodies as seen from a distance of 33 light years. The sun is represented by a blackbody of 5750 K. The planets Jupiter, Venus, Earth, and Mars are shown and are labeled with their first initial. The planets have two peaks in their spectra. The short-wavelength peak is due to sunlight scattered from the planet atmosphere and the long-wavelength peak is from the planet's thermal emission and is estimated by a blackbody of the planet's effective temperature. The high contrast at all wavelengths of the planet-to-star flux ratio is the primary challenge for directly detecting extrasolar planets.*

**Detection of Extrasolar Planets**

The search for extrasolar planets has so far focused on solar system analogues, namely searches for planets orbiting stars like our sun. But, there are many other types of stars that must harbor planets: stars that are much younger, or larger and hotter, or smaller and cooler than the sun. In fact, stars in the longest, main stage of their lives can be up to approximately 100 times more massive, 15 times larger, and 40,000 K hotter than the sun and down to 5 percent of the mass, 10 percent of the radius, and 2500 K less hot than the sun. Terrestrial planets orbiting stars different from the sun would have evolved in very different radiation environments compared to the solar system terrestrial planets.

Despite the sun-like star planet search focus, the first discovered extrasolar planet, in 1992 [4], orbits a distant star extremely different from the sun: a pulsar. A pulsar is the end result of a massive star after a violent "supernova" explosion: a highly compressed star made of neutrons that emits radio waves. Two Earth-mass and one moon-mass planet (together with a giant planet) are known to orbit a pulsar. These planets are not Earth-like; they are very cold and likely inhospitable to life due to deadly radiation emitted by the pulsar. The planets are also too distant to for any future follow-up observations.

Although technology is not advanced enough to detect Earth-mass planets around stars like our sun, 100 extrasolar giant planets have been discovered to orbit nearby, single sun-like stars. The technique used to discover the planets (the "radial velocity method"; see [1,2]) monitors the parent star for evidence of its motion about the common planet-star center of mass, and thus is an indirect detection of a planet. Because only the star is monitored, only the planet's (minimum) mass and orbital parameters can be measured.

The first extrasolar giant planet discovered to orbit a nearby sun-like star [5] shocked the planetary and astronomy communities because the planet was unlike anything in our solar system. This planet, 51 Peg b (named after its parent star 51 Peg A, the 51st brightest star in the constellation Pegasus), is a giant planet with a very short period (4 days) and a correspondingly tiny semi-major axis (0.05 AU)—several times closer to its parent star than Mercury is to our sun. (An AU is astronomical unit, defined as the Earth semi-major axis; 1 AU = 1.5 x $10^{11}$ m.) A giant planet so close to its parent star shook the paradigm of planet formation which had been developed with our own solar system as the only constraint.

The known extrasolar giant planetary systems collectively show that planetary systems can be very different from our own solar system in two important ways. First, the extrasolar giant planets have a surprising range of semi-major axes, distributed almost continuously at different distances from the parent star, up to about 4 AU. Second, many extrasolar planets with a semi-major axis greater than 0.1 AU have eccentric orbits. These two properties are in stark contrast to the solar system where gas giant planets are located beyond 5 AU and all planets have essentially circular orbits. Interestingly, the giant planets also have a very large range of masses, from approximately 0.1 up to (and even beyond) the planetary limit of 13 Jupiter masses.

Where are the solar system analogues with giant planets at 5 AU in circular orbits and potentially unseen terrestrial planets closer to the star? The detection method's selection effects mean we cannot say whether or not solar system analogues are rare. The method favors massive planets close to the star, and can only find planet with periods shorter than the total survey time. The surveys have been going on just under a decade, and would not have been able to detect planets like Jupiter with orbital periods of 12 years. Only 5% of nearby single sun-like stars show evidence for giant planets with orbits < 4 AU [6]. We do not yet know whether these stars, and the other 95% of stars, have distant gas giants such as Jupiter. Over the next five to ten years many planetary systems like our own may be found.

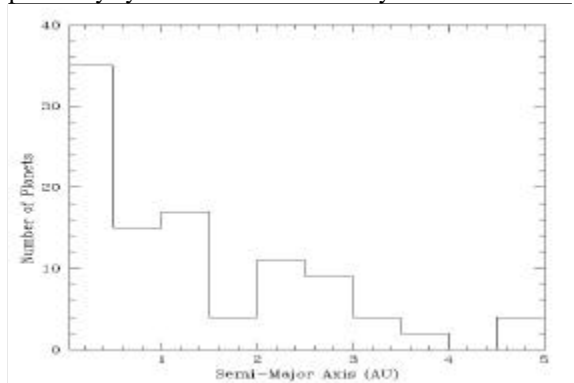

*Fig. 3 A histogram of the semi-major axes of known extrasolar giant planets. The minimum known planet semi-major axis is 0.038 AU. The planets exist at a surprising range of distances from their parent star, and are very different from our own solar system which has gas giants at and beyond 5 AU only. A selection effect makes more massive planets and planets with smaller semi-major axes easier to detect. Data taken from [1].*

The most pressing question for terrestrial planet surveys is *how common are terrestrial-like planets?* Current radial velocity planet searches cannot answer this question, because terrestrial planets orbiting almost all kinds of stars are of too low mass to be detected. A subset of this question answerable with current giant planet searches is *which nearby stars should be searched for Earth-like planets?* The chance of finding Earth-like planets is small around stars with known giant planets near at 0.5-2 AU. These giant planets may have dynamically expelled the much less massive terrestrial planets. Dynamical calculations of the orbital stability of an Earth-mass planet at various semi-major axes for specific systems with known giant planets give a more concrete answer about which stars are dynamically able to harbor terrestrial planets at terrestrial-like semi-major axes.

**Transits as a Planet Detection Method**

A planet transit, shown schematically in Figure 4a, occurs when a planet passes in front of its star as seen from Earth. As the planet crosses the stellar disk, the planet blocks out stellar flux on the ratio of the planet-to-star areas. For an Earth-sized planet transiting a sun-sized star, the drop in the star's brightness would be a miniscule 8 x $10^{-5}$. This is a very small number for astronomical measurements and can only be measured from space, away from the blurring effects of Earth's atmosphere.

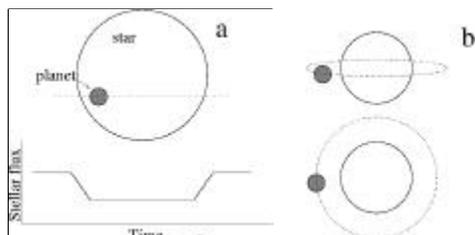

*Fig. 4 Panel a: Schematic diagram of a planet transit. When a planet (represented by the small, dark disk) passes in front of its parent star, the stellar flux (lower part of panel a) drops by the ratio of the planet-to-star areas. This is the case even though stars cannot be spatially resolved as shown in this diagram. Panel b:An extrasolar planet transit will occur only when the planet-star geometry is favorable. Upper diagram shows such a system. In contrast, the lower diagram shows a planet orbit that will never pass in front of the parent star. In reality geometries anywhere in between these will occur.*

In the solar system, Mercury transits the sun several times per century and Venus transits the sun only once (or twice) every 100 years. For extrasolar planets, the probability to transit depends on the orientation of the planet-star system (see Figure 4b). The geometric probability for a planet-star system to be oriented to show transits is the ratio of the stellar radius to planet semi-major axis. For Earth and the sun this is 0.5%, meaning that 200 stars with Earth-like planets would have to be monitored to detect one transiting system. Furthermore, one would want to see two or three transits to measure and confirm the orbital period.

Planet searches that monitor thousands of stars simultaneously to detect the characteristic drop in brightness of a transiting planet will be able to quantify fundamental factors including the frequency of Earth-sized planet, and the terrestrial planet size distribution. Perhaps terrestrial planets will show as broad a size distribution as giant planets do a mass distribution. NASA's Kepler space mission (launch date 2007) and ESA's Eddington space mission (launch date 2008) will each monitor one field of the sky with approximately 100,000 sun-like stars for 3 full years. The COROT mission (CNES and European partners) will be launched in 2005 to detect short-period planets slightly larger than Earth-sized. Many ground-based programs are ongoing to detect short-period giant planets; the lessons learned about false positives will aid the future searches for terrestrial transiting planets.

In the meantime, before the transit searches for Earth-sized planets are launched, the transit method is proving fruitful for extrasolar giant planets. In November 1999 one of the known short-period extrasolar giant planets, HD209458b, was found to transit its parent star [7,8] (see Figure 5 and [9]). The transit detection was a landmark discovery for two important reasons. First, the transit detection confirmed that the then few indirectly detected planet candidates were in fact planets and not some obscure property of the star itself. Second for the first time (and only time to date) an extrasolar planet's radius was measured (at 1.35 $R_{Jupiter}$ [9])—confirming the planet to be a gas giant rather than a very massive rocky planet. Furthermore, the planet size roughly agrees with the theoretical predictions showing that our physical understanding of the planets is generally correct (but cf [10]).

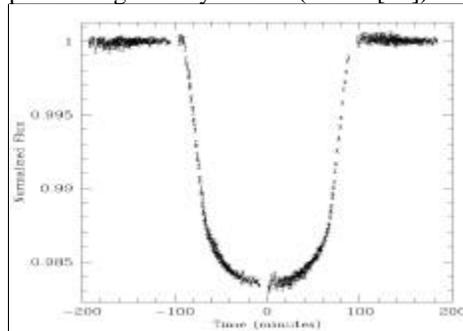

*Fig. 5 Hubble Space Telescope photometry of the one known transiting extrasolar planet [9]. The normalized flux from the parent star drops by the ratio of the planet-to-star areas. The round shape of the transit light curve is due to a non-uniform luminosity across the stellar disk. Data courtesy of D. Charbonneau. At a precision of $10^4$ this is the best measurement of its kind ever. An order of magnitude better must be achieved to detect Earth-sized planets transiting sun-sized stars*

Many follow-up observations of the transiting planet proved to be extremely valuable. In particular *the first extrasolar planet atmosphere detection* was announced in November 2001 when the Hubble Space Telescope observed the "transmission" spectrum of the transiting extrasolar planet HD209458b [11]. As the planet passes in front of its parent star some of the starlight will travel through the planet atmosphere. The planet atmosphere blocks starlight at wavelengths where there are strong absorbers. Hence some of the planetary spectral features are superimposed on the star's spectrum. Specifically absorption from the trace element neutral sodium was detected. The detection of sodium is, after the planet size measurement, a second reassurance for our understanding of the physics of these planets. The absorption feature found was roughly comparable with the theoretical predictions [e.g., 12], even though the conditions in the atmospheres of these planets are very different from those of the solar system giant planets. For example, the transiting planet has an effective temperature of approximately 1100K, compared to Jupiter's 120K, being 100 times closer to its star than Jupiter is to our sun.

**Searching for Earth-like Planets by their Gravitational Influence on the Parent Stars**

*Searches for Earth-mass planets around non-solar-type stars are underway* [13] for short-period Earth-mass planets around the most common type of star: low-mass stars with masses of 0.06 to 0.5 times the mass of the sun. The same technique so successfully used to find the extrasolar giant planets is being used; Earth-mass planets will cause a much larger gravitational effect on low-mass stars than on stars of solar mass. Although common, these low-mass stars are faint and so require difficult and time-consuming observations. Their faintness is because of their small sizes and their low luminosities, corresponding roughly to a blackbody of 3000 K to 4000 K. This low luminosity has two consequences for planets, compared to planets orbiting our sun. The first is that planets will receive a very different wavelength-distribution of starlight, one that peaks at near-infrared wavelengths. The second is that for the same planet-star distance a planet will receive a much lower intensity of starlight. This has an interesting implication—planets with a similar surface temperature and mass to Earth could be found at a much smaller distance to the star. These may be the first Earth-like planets ever detected. In addition, the giant planet searches have found hints of short-period (< 1 month) "super Earths" with masses a few dozen times that of Earth [14] around sun-like stars.

*Near future plans to indirectly detect terrestrial planets* by their gravitational influence on the parent star are focusing on Earth-like planets orbiting sun-like stars at roughly an Earth-sun distance. Just as for the transit surveys described in the above section, the motivation is to determine the frequency of Earth-mass planets around nearby stars and also their mass distribution. Observations from space are required because the Earth's turbulent atmosphere smears starlight and prevents the observational precision required from being reached. For example, for an Earth-mass planet in an Earth-like orbit around a sun-mass star 33 light years away, the star's motion on the sky (as a result of the star's motion about the planet-star common center of mass) would be approximately $10^{-10}$ of a degree. NASA's Space Interferometry Mission (SIM) [15] will be launched in 2009 to search 200 nearby stars, including 50 sun-like stars, for planets down to several Earth-masses. One challenge for SIM is that in a multiple planet system giant planets will dominate the star's astrometric signature. For a description of this and other methods to detect Earth-mass planets by their gravitational influence on the parent star see [15,16,17].

**Direct Planet Detection Developments**

Direct detection of reflected or emitted light from the planet itself is the only method that will allow study of planet atmospheres and surface characteristics. Direct detection is extremely difficult because of the light from the parent star scattered in the telescope optics makes it almost impossible to detect any object close to the star (see Figure 6). Even with direct detection the planets will be observed as a point source of flux. In other words there will be no spatial resolution of the planetary surface or atmosphere.

The enormous brightness contrast between an Earth-like planet and its parent star—6 to 10 orders of magnitude (see Figure 2)—presents an immense technological challenge for direct detection: to block out the light of the parent star. In addition, the telescope must be large enough to observe the planet point source separately from the stellar point source. These combined requirements, "starlight rejection" and "resolution" results in a very different design at different wavelengths.

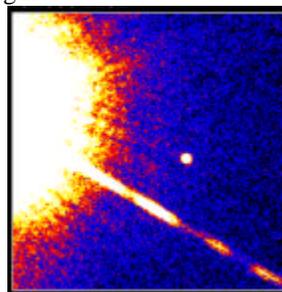

*Fig.6 Hubble Space Telescope image of the star Gliese 229 (left) and its much fainter substellar companion (right, small object). The primary star is so bright compared to the companion that it floods the telescope detector making it look very different from a point source. The diagonal line is a diffraction spike produced by scattered starlight in the telescope's optical system. An Earth-like planet would be 50 times closer to the primary star center and 1,000,000 times fainter—far from detectable with this set up. Credit: S. Kulkarni, D. Golimowski, and NASA.*

Two kinds of direct imaging techniques are being developed, one for visible wavelengths (0.5 to 1 micron) and another for mid-IR wavelengths (7 to 20 micron). A spectral resolution (defined by $\lambda/\Delta\lambda$ where $\lambda$ is wavelength) of 20 is considered sufficient, and observing times of hours to detect terrestrial planets and up to weeks to obtain atmospheric spectra is considered acceptable. Although future 30-m or 100-m ground-based

6telescopes may be able to directly detect Earth-like planets (e.g., [18]), concept and technology development so far has been for space-based telescopes. Both NASA and ESA are developing technology for space mission direct detection of terrestrial-like planets around stars within 300 light years, despite the fact that it is a > 1 billion dollar cost and will have launch dates in the 2015 time frame.

ESA is developing "Darwin" a mid-infrared wavelength space interferometer [16,19]. The Darwin design calls for 6 telescopes each at least 1.5 m in diameter. The telescopes will be arranged in a hexagon shape with a baseline of 10s to 100s of meters and with a central beam-combining unit. The telescopes will be used as an interferometer where the combination of signals from individual telescopes will mimic a much larger telescope, needed to separate the planet and star. The interferometer will be used to "null" out the starlight using destructive interference by a delay in signals from some telescopes. In order for nulling interferometry to succeed, the distances between the telescopes must be precisely controlled—to the centimeter level. Hence formation flying control and precise path length control are two of the most serious technical challenges to the "free-flying nulling interferometer".

NASA is considering a very similar design to Darwin for its "Terrestrial Planet Finder" (TPF); a preliminary design is described in [20] and updates are on [15]. A recent, dramatic new development lead to consideration of a second architecture concept: a single telescope to operate at visible wavelengths much like a follow-on to the 2 meter Hubble Space Telescope. One technological benefit is that at visible wavelengths a single, ~8 m mirror telescope will give a similar resolution as a multi-telescope mid-IR wavelength free flying interferometer. The initial problem arising from the huge planet-star brightness ratio is diffracted light around the telescope aperture; this "noise" prevents any object being detected near the star (see Figure 6). The new development is to drastically modify the effective shape of the telescope aperture by using a specially shaped mask (or "pupil") in the optical path. (This can be thought of as changing the telescope mirror shape.) This pupil will use the diffraction of light around the pupil edges for an advantage: to minimize the diffracted light at some locations in the image. This shaped pupil (see [21] and references therein) can in principle block up to 10 orders of magnitude of starlight although current lab test beds are not this sensitive. In addition to the shaped pupil, a "classical coronagraph", a traditionally used circular device to block out starlight in the image plane, could be used. The two most serious technological challenges to the single-mirror visible-wavelength telescope are the diffracted light problem and also scattered light from miniscule (0.1 nm) imperfections in the mirror surfaces. Both of these problems are at a level of a million times too high for conventional telescopes. NASA is studying both the free flying nulling interferometer and the visible wavelength telescope and will decide which technology to pursue for TPF in 2006.

**Remote Sensing for Planetary Biomarkers**

The search for Earth-like planets includes the goal of studying the planet atmosphere for evidence of habitability and life. The idea of looking at an atmosphere for indications of life was first addressed by Lederberg [22] and Lovelock [23] who suggested searching for signs of severe chemical disequilibrium. With Earth as our only example of a planet whose atmosphere is modified by life, it is clearly the best place to start for specific examples of atmospheric biomarkers.

There are several strong spectral diagnostics in Earth's atmosphere for habitability and life as we know it (Figures 7 and 8). These spectral features are described in detail in [24]. $O_2$ and its photolytic product $O_3$ are the most reliable biomarker gas indicators for life as we know it. $O_2$ is highly reactive and therefore will remain in significant quantities in the atmosphere only if it is continually produced. There seem to be no non-biological sources that can continually produce large quantities of $O_2$ (Venus and Mars both have very small amounts of $O_2$) and only rare false positives that in most cases could likely be ruled out by other planetary characteristics. $O_3$ is a non-linear indicator of $O_2$; only a small amount of $O_2$ need be present to produce a relatively large quantity of $O_3$. $N_2O$ is a second gas produced by life—albeit in small quantities—during microbial oxidation-reduction reactions. $N_2O$ has a very weak spectroscopic signature. Furthermore, because the $N_2O$ spectroscopic signature overlaps with $H_2O$ absorption bands, it may be more easily detectable on a planet with less water vapor than Earth. $H_2O$ vapor itself is a very important spectral feature. $H_2O$, while not a biomarker, is indicative of habitability because all known life needs liquid water. Other spectral features, while not necessarily biomarkers or habitability indicators, can provide useful planet characterization information. $CO_2$ is indicative of a terrestrial planet atmosphere; Venus, Earth, and Mars all have $CO_2$ which is easily detectable due to a very strong mid-IR absorption feature. $CO_2$ originates from outgassing and is expected to be present on all small, rocky planets. High concentrations of $CH_4$, such as may have been present on early Earth, could indicate the presence of methanogenic bacteria, but could also be from midocean ridge volcanism. Other gases do not have strong enough absorption features (even at higher abundance than Earth levels) to be detectable with the expected signal-to-noise and spectral resolution. The simultaneous detection of combinations of the above features may be a more robust indicator of life or habitability than detection of only a single unusual feature (e.g., [25]). In relation to this there is some debate as to which wavelength regime is most suitable for life and habitability indicators.

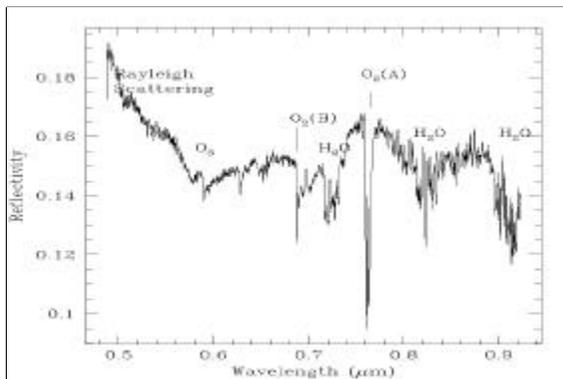

*Fig.7 A visible wavelength spectrum of the spatially unresolved Earth, as seen with Earthshine [26]. The viewpoint is largely centered equatorially on the Pacific Ocean. The major atmospheric features are identified. The reflectivity scale is arbitrary. Data courtesy of N. Woolf and W. Traub.*

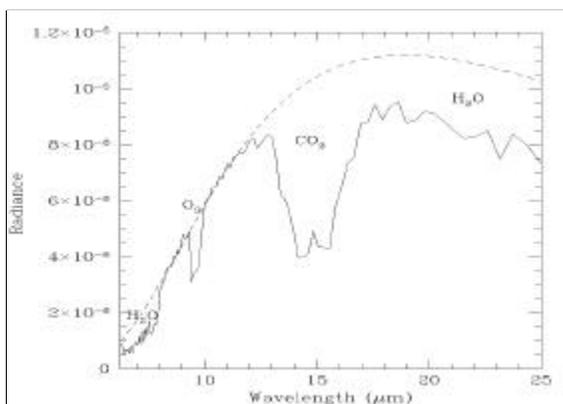

*Fig.8 A mid-infrared wavelength spectrum of the spatially unresolved Earth taken by the Thermal Emission Spectrometer experiment on the Mars Global Surveyor spacecraft en route to Mars in 1996 [35]. The viewpoint is over Hawaii. The structure in the water vapor bands is due to many strong, widely spaced spectral lines. Note also the emission at the center of the $CO_2$ band. Radiance is in units $W\,cm^{-2}\,sr^{-1}\,cm^{-1}$.*

In addition to atmospheric biomarkers Earth has one very strong and very intriguing biomarker on its surface: vegetation. The reflection spectrum of photosynthetic vegetation has a dramatic sudden rise in albedo around 750 nm by a factor of five or more. This "red-edge" feature is caused both by strong chlorophyll absorption to the blue of 700 nm and a high reflectance due to plant cell structure to the red of 700 nm. If plants did not have this window of strong reflection (and transmission), they would be become too warm and their chlorophyll would degrade. On Earth, this red-edge signature is probably reduced to a few percent [26,27] due largely to cloud cover and noncontinuous vegetation cover across the planet. Such a spectral surface feature could be much stronger on a planet with a lower cloud cover fraction. Recall that any observations of extrasolar Earth-like planets will not be able to spatially resolve the surface. A surface biomarker could be distinguished from an atmospheric signature by time variation; as the continents, or different concentrations of the surface biomarker, rotate in and out of view the spectral signal will change correspondingly.

We should not be held hostage to our only example of planetary life. Indeed Earth has exhibited very different atmospheric signatures in the past due to extreme climatic states such as glaciation, the rise of photosynthetic organisms, and the methane bursts that may have punctuated Earth's history [28]. Moreover, just as each of the solar system terrestrial planets differ greatly from each other, there is no reason to expect extrasolar Earth-like planets to be similar to Earth. If an Earth-like planet atmosphere is determined to have a severe departure from chemical and thermodynamic equilibrium would we be able to identify the disequilibrium features with biological modification of atmospheric composition? Or will there always be an ambiguity with geological processes? Will we be able to unequivocally identify a spectral signature not consistent with any known atomic, molecular, or mineral signature in the solar system and Universe (such as Earth's vegetation red edge)? Preparations to try to answer these questions are underway with theoretical models spanning a huge range of parameter space (e.g., [25,29]). Earth-like planet atmospheres can be computed using known atmospheric physics and chemistry under a variety of conditions and inputs. If all combinations of spectral features can be identified with known atmospheric physics and chemistry we may be able to identify any unusual signatures possibly attributable to life.

**Studying Earth as an Extrasolar Planet**

We know surprisingly little about what information could be extracted if Earth were a distant extrasolar planet observable only as a point source of light. An emerging field of research aims to understand Earth as an extrasolar planet both with models and with observations.

Recent modeling work [30] has shown that physical properties of Earth may be extracted by monitoring the time variation of the visible flux of Earth. Cloud patterns due to meteorological activity together with water vapor in the atmosphere would be indicative of large bodies of liquid water. As Earth rotates and continents come in and out of view, the total amount of reflected sunlight will change due to the high albedo contrast of different components of Earth's surface (<10% for ocean, >30-40% for land, >60% for snow and some types of ice). In the absence of clouds this variation could be an easily detectable factor of a few. With clouds the variation is muted to 10 to 20%. Nevertheless, the rotational period could be extracted. From a planet with much less cloud cover than Earth, much more surface information may be extracted.

Real data of the spatially unresolved Earth is available. Global, instantaneous spectra and photometry can be obtained from observations from Earth itself—by Earthshine measurements. Earthshine, easily seen with the naked eye during crescent moon phase (Figure 9), is sunlight scattered from Earth that scatters off of the moon and travels back to Earth. Earthshine data is more relevant to studying Earth as an extrasolar planet than remote sensing satellite data. The latter is highly spatially resolved and limited to narrow spectral regions. Furthermore by looking straight down at specific regions of Earth, hemispherical flux integration with lines-of-sight through different atmospheric path lengths is not available.

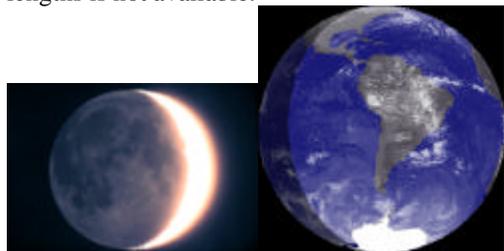

*Fig. 9 A photograph of Earthshine (left image) courtesy of A. Wong [31] and a view of the Earth as seen from the moon (right image) [32]. Earthshine is scattered sunlight from Earth that illuminates the dark part of the moon. The moon and Earth always have opposite phases.*

Several Earthshine studies are ongoing including one [33] to monitor changes in Earth's global albedo as a proxy for climate change. More recent efforts have focused on a broad wavelength coverage of spectra [26,27] and have shown: the Earth's blue color from Rayleigh scattering; the major absorption features at visible wavelengths; and possibly the Earth's vegetation red edge signature. Note that the Earth's surface can be studied only where the atmosphere is transparent, at visible wavelengths and at mid-IR wavelengths in the 8—12 micron window.

Turning space satellites to look at Earth is also very useful because ultimately we need to consider viewing geometries other than the equatorial viewpoint probed with Earthshine measurements. In a landmark study the Galileo spacecraft took some visible and near-IR spectra of Earth during two gravitational assist flybys of Earth. Looking at small areas of Earth, Sagan et al. [34] concluded that the widespread presence of the vegetation red edge combined with abundant atmospheric oxygen, other molecules out of thermodynamic equilibrium and radio signals "constituted evidence of life on Earth without any *a priori* assumptions about its chemistry". Later a spatially unresolved mid-IR spectrum of Earth was taken with Thermal Emission Spectrometer experiment on the Mars Global Surveyor spacecraft en route to Mars [35] (Figure 8). Even more useful would be a time series of spatially unresolved spectra of Earth at different viewing geometries including polar. A small satellite for this purpose has been studied by the Canadian Space agency [36].

**Future Prospects**

The field of extrasolar planet research began in earnest in 1995 and since then has undergone an explosion. Much work needs to be done to detect planets like Earth with similar temperatures and mass orbiting nearby stars. The earliest detection could be within a few years if very low-mass stars have a large frequency of Earth-like planets with small planet-star distances. Otherwise the detection of Earth-like planets will have to wait until space missions are launched or large ground-based telescopes (30 to 100 m) are built within the next two decades.

Towards the end of this decade, returns from space missions to indirectly detect Earth-like planets will tell us the frequency, the size distribution, and the mass distribution of terrestrial planets. This data will provide invaluable constraints for understanding planet formation. During the latter part of the second decade from now, if space missions proceed as planned, Earth-like planets will be directly detected around nearby stars. With direct detection (and depending on the wavelength capability of the space mission) we can measure the planet's reflected flux or thermally emitted flux, the atmospheric spectrum, and possibly the planet's rotational period and very gross surface characteristics. If terrestrial planets are common we could have dozens of them with the above properties to compare with Earth—the search for extrasolar Earth-like planets will move from astronomy into crude comparative planetology.

In the meantime, many exciting discoveries await us and there is much work to be done with extrasolar giant planets. Giant planets are stepping stones to detecting and characterizing Earth-like planets through developing new techniques that will be used later for Earth-like planets, using the known giant planetary systems to constrain models of planet formation, and understanding the atmospheres of planets in new environments. It is not unrealistic to expect the discovery of hundreds of extrasolar planets in the next few decades.

When an Earth-like planet orbiting a sun-like star with an unusual, highly disequilibrium atmospheric signature is detected, the urge to press for more information will be irresistible. Very long term plans are being thought out for enormous space telescopes to study the atmospheres and in greater detail for signs of life [37] and to study the planet surfaces by planet images with many pixels by using a very large number of space telescopes [38], and even to send space craft to such planets [39].

**Acknowledgements**

I am extremely grateful to John Bahcall for valuable discussions. I would like to thank the members of the Princeton Terrestrial Planet Finder group for very stimulating discussions over the past two years. I also thank Mike Wevrick, Wes Traub, and Lisa Kaltenegger for useful discussions. I thank the referees Marc Kuchner, Jim Kasting, and an anonymous referee for helpful reviews. This work was supported by the W. M. Keck Foundation and by the Carnegie Institution of Washington.

**References**

1. The California planet search: http://exoplanets.org/
2. Geneva Planet Search: http://obswww.unige.ch/~udry/planet/elodie.html
3. The Extrasolar Planet Encyclopaedia: http://www.obspm.fr/encycl/encycl.html
4. A. Wolszczan, D. Frail, D., A planetary system around the millisecond pulsar PSR1257+12, Nature 255 (1992) 145-147.
5. M. Mayor, D. Queloz, A Jupiter-mass companion to a solar-type star, Nature 378 (1995) 355-359.
6. R.P. Butler, G.W. Marcy, D.A. Fischer, S.S. Vogt, C.G. Tinney, H.R. Jones, A.J. Penny, K. Apps, in Planetary Systems in the Universe: Observation, Formation and Evolution, IAU Symp. 202, Eds. A. Penny, P. Artymowicz, A.-M. Lagrange and S. Russel ASP Conf. Ser., in press.
7. D. Charbonneau, T.M. Brown, D.W. Latham, M. Mayor, Detection of planetary transits across a sun-like star, ApJ 529 (2000) L45-48.
8. G.W. Henry, G.W. Marcy, R.P. Butler, S.S. Vogt, A Transiting "51 Peg-like" planet, ApJ 529 (2000) L41-44.
9. T. M. Brown, D. Charbonneau, R.L. Gilliland, R.W. Noyes, A. Burrows, Hubble Space Telescope Time-series photometry of the transiting planet of HD 209458, ApJ 552 (2001) 699-709.
10. T. Guillot, A.P. Showman, Evolution of "51 Pegasus b-like" planets, A&A 385 (2002) 156-165
11. D. Charbonneau, T. M. Brown, R.W. Noyes, R.L. Gilliland, Detection of an extrasolar planet atmosphere, ApJ 568 (2002) 377-384.
12. S. Seager and D.D. Sasselov, Theoretical transmission spectra during extrasolar giant planet transits, ApJ 537 (2000) 916-921.
13. M. Endl, M. Kurster, F. Rouesnel, S. Els, A.P. Hatzes, W.D. Cochran, Extrasolar terrestrial Planets: can we detect them already?, in ASP Conf. Ser., Scientific Frontiers in Research on Extrasolar Planets, ed. D. Deming and S. Seager (San Francisco: ASP) 294 (2003) 75-78.
14. G.W. Marcy, private communication.
15. http://planetquest.jpl.nasa.gov/
16. http://sci.esa.int/
17. W.A. Traub, K.A. Jucks, A possible aeronomy of extrasolar terrestrial planets, in Atmospheres in the solar system: comparitive aeronomy, eds M. Medilo, A. Nagy, H.J. Waite, AGU Geophysical Monograph 130 (2002) 369-380
18. J.R.P. Angel, Imaging terrestrial exoplanets from the ground, in ASP Conf. Ser., Scientific Frontiers in Research on Extrasolar Planets, ed. D. Deming and S. Seager (San Francisco: ASP) 294 (2003) 543-556.
19. The European Space Agency, Darwin the infrared space interferometer: concept and feasibility study report, ESA-SCI (2000) 12
20. C.A. Beichman, N.J. Woolf, C. A. Lindensmith, The Terrestrial Planet Finder (TPF), NASA/JPL Publication 99-3.
21. M.J. Kuchner, W.A. Traub, A coronagraph with a band-limited mask for finding terrestrial planets, ApJ 570 (2002), 900-908.
22. J. Lederberg, Signs of life: criterion-system of exobiology, Nature 207 (1965) 9-13.

23. J.E. Lovelock, A physical basis for life detection experiments, Nature 207 (1965) 568-570.
24. D. J. Des Marais, M. Harwit, K. Jucks, J. Kasting, D. Lin, J. Lunine, J. Schneider, S. Seager, W. Traub and N. Woolf, Remote sensing of planetary properties and biosignatures on extrasolar terrestrial planets, Astrobiology, 2, 153-181 (2002).
25. F. Selsis, D. Despois, J.-P. Parisot, Signature of life on exoplanets: can Darwin produce false positive detections?, A&A 388 (2002) 985-1003..
26. N.J. Woolf, P.S. Smith, W.A. Traub, K.W. Jucks, The spectrum of earthshine: a pale blue dot observed from the ground, ApJ 574 (2002) 430-433.
27. L. Arnold, S. Gillet, O. Lardière, P. Riaud, J. Schneider, A test for the search for life on extrasolar planets: Looking for the terrestrial vegetation signature in the Earthshine spectrum, A&A 392 (2002) 231-237.
28. S. Bains, R.M. Corfield, R.D. Norris, Mechanisms of climate warming at the end of the Paleocene, Science 285 (1999) 724-727.
29. V.S. Meadows et al., The virtual planetary laboratory: towards characterizing extrasolar terrestrial planets, AAS DPS meeting #33, #40.12 2001.
30. E.B. Ford, S. Seager, E.L. Turner, Characterization of extrasolar terrestrial planets from diurnal photometric variability, Nature 412 (2001) 885-887.
31. http://www.home.earthlink.net/~alsonwong/earthshi.htm
32. http://www.fourmilab.ch/earthview/
33. P.R. Goode, J. Qui, V. Yurchyshyn, J. Hickey, M-C Chu, E. Kolbe, C.T. Brown, S.E. Koonin, Earthshine observations of the Earth's reflectance, J. Geophys. Res. Lett, 28 (2001) 1671-1674.
34. C. Sagan, W.R. Thompson, R. Carlson, D. Gurnett, C.A. Hord, Search for life on Earth from the Galileo spacecraft, Nature 365 (1993) 715-721.
35. J.C. Pearl, P.R. Christensen, Initial data from the Mars Global Surveyor thermal emission spectrometer experiment: Observations of the Earth, JGR 102 (1997) 10875-10880.
36. G.R. Davis, S.B. Calcutt, J.R. Drummond, D.A. Naylor, A.J. Penny, S. Seager, Measurements of the Unresolved Spectrum of Earth (MUSE), Final Report of the Concept Study for the Canadian Space Agency Space Science Program, April 2002.
37. http://origins.jpl.nasa.gov/missions/lf.html
38. http://origins.jpl.nasa.gov/missions/pi.html
39. S. Kilston, THE PROJECT: an Observatory / Transport Spaceship for Discovering and Populating Habitable Extrasolar Terrestrial Planets, BAAS 30 (1998) 1392.

Sara Seager is a research staff member at the Department of Terrestrial Magnetism of the Carnegie Institution of Washington, located in Washington, DC. Her theoretical work focuses on modeling extrasolar planet atmospheres to predict observational signatures and to interpret data. She is also co-leading a search for extrasolar transiting planets. Seager received a Ph.D. in astronomy from Harvard University in 1999 and then spent three years at the Institute for Advanced Study in Princeton, NJ as a long-term member. She has been at the Carnegie Institution of Washington since August 2002.